\documentclass[lettersize,journal]{IEEEtran}
\IEEEoverridecommandlockouts

\usepackage{amssymb}
\usepackage[cmex10]{amsmath}
\usepackage{stfloats}
\usepackage{graphicx}
\usepackage{subfigure}
\usepackage{tabularx}
\usepackage{epsfig,epsf,color,balance,cite}
\usepackage{verbatim}
\usepackage{url}
\usepackage{bm}
\usepackage{booktabs}
\usepackage{siunitx}

\usepackage{amsmath}
\allowdisplaybreaks[2]
\usepackage{ntheorem}

\usepackage{algorithm}
\usepackage{algorithmic}

\usepackage{color}
\definecolor{myc1}{rgb}{0,0,0}

\begin{document}

\title{WirelessAgent: A Unified Agent Design for General Wireless Resource Allocation Problem without Current Channel State Information}

\author{Ran Yi,
        Ruopeng Xu,
        Dongshu Zhao,
        Zhaoyang Zhang, 
        Baolin Chen,
        Kai-Kit Wong,
        Hyundong Shin,
        Zhaohui Yang

\thanks{Ran Yi, Ruopeng Xu, Zhaohui Yang, and Zhaoyang Zhang are with the College of Information Science and Electronic Engineering, Zhejiang University, and also with Zhejiang Provincial Key Laboratory of Info. Proc., Commun. \& Netw. (IPCAN), Hangzhou 310027, China (e-mails: \{ranyi, ruopengxu, yang\_zhaohui, ning\_ming\}@zju.edu.cn).}
\thanks{Dongshu Zhao is with the School of Art and Archaeology, Zhejiang University, Hangzhou 310058, China (e-mail: dongshu\_zhao@zju.edu.cn).}
\thanks{Baolin Chen, B. Chen is with Zhejiang Institute of Communications Co., LTD., Hangzhou, Zhejiang 310030, China (e-mail: chenbl@zjic.com).}
\thanks{Hyundong Shin,  H. Shin is with the Department of Electronics and Information
 Convergence Engineering, Kyung Hee University, Yongin-si, Gyeonggi-do
 17104, Republic of Korea (e-mail: hshin@khu.ac.kr).}
\thanks{Kai-Kit Wong,  K. Wong is affiliated with the Department of Electronic and ElectricalEngineering, University College London, Torrington Place, UK and also with
the Department of Electronic Engineering, Kyung Hee University, Yongin-si,Gyeonggi-do 17104, Republic of Korea (e-mail: kai-kit.wong@ucl.ac.uk).}
\vspace{-1em}
}

\maketitle

\begin{abstract}
This paper investigates the agent design for solving the wireless resource allocation problem without sufficient channel state information  (CSI), which cannot be effectively solved via conventional method. In the considered wireless agent design, we provide the general sense-repair-decide-act workflow, which can be used to intelligently solve  general wireless resource allocation problem. A multi-objective optimization problem is formulated to adaptively satisfy different user requirements including both spectrum and energy efficiency. This work addresses the challenge of incomplete CSI for multiple optimization objectives. To solve this problem, we use an artificial intelligence (AI) model to predict missing channel data and construct an agent on the Coze platform, allowing the network operators to optimize multiple objectives through natural language conversations. To tackle the resource scheduling under different objectives, we develop adaptive algorithms. Simulation results validate the effectiveness of our proposed design, demonstrating that the proposed AI method reduces the root mean square error by approximately up to 67\% compared to the traditional approach. Moreover, the data-driven scheduling balances system performance compared to conventional baseline approaches.
\end{abstract}


\IEEEpeerreviewmaketitle

\section{Introduction}
\label{sec:intro}

\IEEEPARstart{D}{riven} by the growth of mobile data traffic and the ambitious sustainability targets of sixth generation (6G) networks, orthogonal frequency division multiple access (OFDMA) has established itself as a fundamental technology for achieving high spectral efficiency and massive connectivity \cite{9600846}. In OFDAM-enabled systems\cite{9296440}, dynamic resource allocation, including subcarrier scheduling and power control, plays a pivotal role in optimizing network performance\cite{10144244}. While the conventional mathematical optimization theory based algorithms are critical, their actual use faces two challenges: the heavy reliance on perfect channel state information (CSI) which is often impractical, and the increasing operational complexity of manually choosing these diverse algorithms to meet different user requirements. The integration of agentic artificial intelligence (AI) platforms into communication system that can autonomously manage data inaccuracy and algorithm selection via natural language interaction has emerged as a critical enabler for future intelligent 6G network management.

However, in practical deployment scenarios, acquiring perfect CSI is a significant challenge. The channel data matrix is often sparse or incomplete due to various factors such as feedback link congestion, hardware failures, or rapid time-varying fading \cite{10275111}. Traditional methods\cite{10915662,xu2025energy} for handling resource allocation with partial CSI, such as zero-filling or statistical mean imputation, fail to capture the correlations of wireless channels in the frequency and spatial domains, leading to suboptimal resource allocation and performance degradation. Furthermore, unlike traditional networks that optimize for a single static metric (e.g., sum rate), future 6G applications demand a high degree of flexibility to satisfy diverse user requirements, from ultra-high throughput for video streaming to minimizing power consumption for Interent of things (IoT) networks \cite{9955312}.

While advanced channel estimation techniques and resource allocation algorithms have been extensively studied\cite{9921202,10032275}, the research on how to maintain system performance without current under imperfect CSI while adapting to dynamic user objectives remains a gap. Prior works\cite{9921202,10032275} have generally focused on either deep learning-based channel estimation \cite{9921202} or convex optimization for a specific resource scheduling problem \cite{10915662}, without jointly considering the interplay between data availability and intent-driven multi-objective optimization. For instance, robust scheduling algorithms\cite{9921202} often assume a bounded error model rather without using the history information, while AI-based channel predictors rarely integrate seamlessly with downstream control logic for energy efficiency (EE) or power minimization trade-offs.

In this paper, we propose an agent-based general resource allocation framework for OFDMA systems. We consider a scenario where the base station (BS) must schedule resources under the constraint of incomplete CSI and varying service requirements. 

The main contributions of this work are summarized as follows:
\begin{itemize}
    \item We develop an agentic AI framework on the Coze platform that autonomously operates a ``Sense-Repair-Decide-Act'' workflow. By enabling natural language interaction, this agent addresses the operational inefficiency of manually configuring diverse optimization algorithms, realizing seamless switching among complex objectives.
    \item To ensure robustness against incomplete CSI, we integrate a nearest-neighbor collaborative filtering neural network compression framework (NNCF) algorithm within the agent. This module accurately recovers missing channel data by exploiting frequency-spatial correlations, providing higher-quality input to make decisions.
    \item We formulate a multi-objective optimization problem that adaptively targets at maximum rate, minimum power, or maximum EE based on recognized user requirements. Efficient algorithms, including water-filling\cite{9921202}, channel inversion\cite{9921202}, and iterative power search\cite{9921202}, are developed to solve these problems, with simulation results suggesting additional performance gains over conventional baselines.
\end{itemize}

\section{System Model and Problem Formulation}
\begin{figure*}[!t] 
    \centering
    \includegraphics[width=1\textwidth, trim=0.0cm 11.2cm 0.0cm 3.0cm, clip]{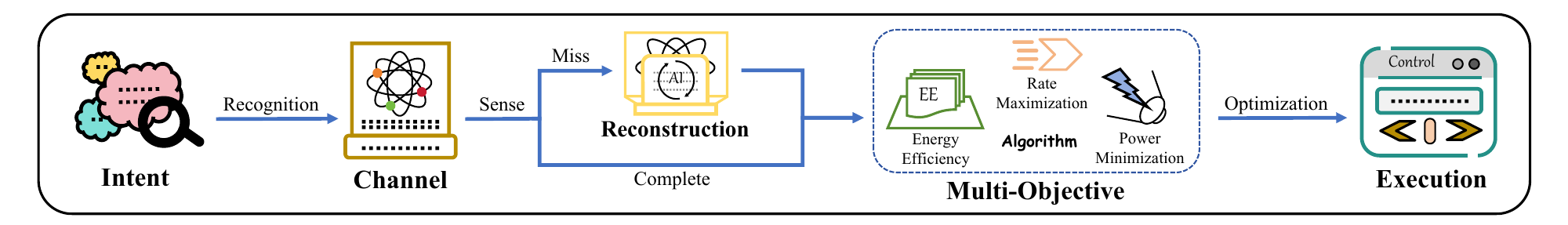}
    \caption{The proposed agent-enabled general resource allocation framework. The workflow includes intent recognition, AI-based channel reconstruction for missing CSI, and multi-objective adaptive optimization (max rate, min power, max EE) to generate optimal execution commands.}
    \label{fig:framework}
\end{figure*}
\label{sec:system_model}

\subsection{Network Model}
Consider a downlink OFDMA wireless communication system consisting of a single BS and $K$ single-antenna users, denoted by the set $\mathcal{K} = \{1, 2, \dots, K\}$. The total system bandwidth $B_{\mathrm{total}}$ is divided into $N$ orthogonal subcarriers, denoted by the set $\mathcal{N} = \{1, 2, \dots, N\}$, each with a bandwidth of $B_{\mathrm{total}}$.

The CSI is critical for resource allocation. Let $g_{k,n}$ denote the channel power gain between the BS and user $k$ on subcarrier $n$. In an ideal scenario, the BS requires the complete channel matrix $\mathbf{G} \in \mathbb{R}^{K \times N}$. The achievable data rate of user $k$ on subcarrier $n$ can be expressed as:
\begin{equation}
    r_{k,n} = B \log_2 \left( 1 + \frac{p_{k,n} g_{k,n}}{N_0 B} \right),
    \label{eq:shannon}
\end{equation}
where $p_{k,n}$ is the transmit power allocated to user $k$ on subcarrier $n$, and $N_0$ is the power spectral density of the additive white Gaussian noise (AWGN).

\subsection{Channel Missing and AI-Based Repair Model}
In practical deployments, acquiring perfect CSI is challenging due to feedback delay, hardware limitations, or environmental interference. We assume that the BS initially obtains an incomplete raw channel matrix, denoted by $\mathbf{G}_{raw}$. The relationship between the observed and true channel gains can be modeled as:
\begin{equation}
    [\mathbf{G}_{raw}]_{k,n} = 
    \begin{cases} 
    g_{k,n}, & \text{if } (k,n) \in \Omega \\
    \text{NaN}, & \text{otherwise},
    \end{cases}
\end{equation}
where $\Omega$ represents the set of indices for successfully received CSI, and $\text{NaN}$ denotes missing data.

To handle the impact of imperfect CSI, we introduce an AI-based channel reconstruction module. By using the correlations of wireless channels in both the frequency domain and the spatial domain, the module reconstructs the missing entries. The repaired complete channel matrix is denoted by $\hat{\mathbf{G}}$. All the following resource allocation decisions are based on $\hat{\mathbf{G}}$.

\subsection{Problem Formulation}
I introduce a binary subcarrier allocation variable $x_{k,n} \in \{0, 1\}$. $x_{k,n} = 1$ indicates that subcarrier $n$ is allocated to user $k$, and $x_{k,n} = 0$ otherwise. In OFDMA, each subcarrier is allocated to at most one user to avoid co-channel interference:
\begin{equation}
    \sum_{k=1}^{K} x_{k,n} \le 1, \quad \forall n \in \mathcal{N}^{\cdot}
    \label{eq:ofdma_constraint}
\end{equation}

The system is designed to adaptively switch between different optimization objectives based on user requirement. We formulate three distinct optimization objectives corresponding to maximum rate, minimum power, and maximum EE.

\subsubsection{Maximum Rate}
When high throughput is prioritized (e.g., video streaming), the objective is to maximize the system sum rate subject to a total power budget $P_{max}$. Let $\mathbf{X} = \{x_{k,n} \mid k \in \mathcal{K}, n \in \mathcal{N}\}$ and $\mathbf{P} = \{p_{k,n} \mid k \in \mathcal{K}, n \in \mathcal{N}\}$ denote the sets of subcarrier allocation variables and power allocation variables for all users and subcarriers. The problem is formulated as:
\begin{subequations}
\begin{align}
    \textbf{P1:} \quad \max_{\mathbf{X}, \mathbf{P}} \quad & \sum_{k=1}^{K} \sum_{n=1}^{N} x_{k,n} r_{k,n} \\
    \text{s.t.} \quad & \sum_{k=1}^{K} \sum_{n=1}^{N} x_{k,n} p_{k,n} \le P_{max}, \label{eq:p1_power} \\
    & p_{k,n} \ge 0, \quad \forall k, n, \\
    &\sum_{k=1}^{K} x_{k,n} \le 1, \quad \forall n \in \mathcal{N}^{\cdot}, \quad x_{k,n} \in \{0, 1\}.
\end{align}
\end{subequations}

\subsubsection{Minimum Power}
For energy-constrained scenarios (e.g., Interent of things networks), the objective is to minimize total transmit power while satisfying a minimum data rate requirement $R_{min}$ for each user:
\begin{subequations}
\begin{align}
    \textbf{P2:} \quad \min_{\mathbf{X}, \mathbf{P}} \quad & \sum_{k=1}^{K} \sum_{n=1}^{N} x_{k,n} p_{k,n} \\
    \text{s.t.} \quad & \sum_{n=1}^{N} x_{k,n} r_{k,n} \ge R_{min}, \quad \forall k \in \mathcal{K}, \label{eq:p2_qos} \\
    & p_{k,n} \ge 0, \quad \forall k, n, \\
    &\sum_{k=1}^{K} x_{k,n} \le 1, \quad \forall n \in \mathcal{N}^{\cdot}, \quad x_{k,n} \in \{0, 1\}.
\end{align}
\end{subequations}

\subsubsection{Maximum EE}
To balance performance and EE, the system optimizes energy efficiency, defined as the ratio of total sum rate to total power consumption (including a fixed circuit power $P_c$):
\begin{subequations}
\begin{align}
    \textbf{P3:} \quad \max_{\mathbf{X}, \mathbf{P}} \quad & \frac{\sum_{k=1}^{K} \sum_{n=1}^{N} x_{k,n} r_{k,n}}{\sum_{k=1}^{K} \sum_{n=1}^{N} x_{k,n} p_{k,n} + P_c} \\
    \text{s.t.} \quad & \sum_{k=1}^{K} \sum_{n=1}^{N} x_{k,n} p_{k,n} \le P_{max}, \label{eq:p3_power} \\
    & p_{k,n} \ge 0, \quad \forall k, n, \\
    &\sum_{k=1}^{K} x_{k,n} \le 1, \quad \forall n \in \mathcal{N}^{\cdot}, \quad x_{k,n} \in \{0, 1\}.
\end{align}
\end{subequations}

The formulated problems P1-P3 involving binary variable $x_{k,n}$ and continuous variable $p_{k,n}$ are mixed-integer non-linear programming (MINLP) problems. In the following section, we propose an agentic AI algorithm to solve the channel reconstruction and resource allocation problems sequentially.

\subsection{Agentic AI Implementation on Coze Platform}
\label{sec:coze_agent}

To bridge the gap between complex optimization algorithms and real network management, we use the proposed framework as an agentic AI on the Coze platform. This platform acts as the central engine, enabling natural language interaction and autonomous workflow execution. As illustrated in Fig. \ref{fig:framework}, the overall agent architecture consists of three key layers:


\subsubsection{Perception Layer (Intent Recognition)}
The agent uses the semantic understanding property of large language models (LLMs) to interpret informal user queries (e.g., ``Higher speed for video streaming'' and ``Save power for sensors''). The LLM maps these queries to specific intent flags ($I \in \{\text{max rate, min power, max EE}\}$), eliminating the need for manual parameter configuration.

\subsubsection{Tool Layer (Functional Plugins)}
We map the math algorithms into independent, ready-to-use plugins:
\begin{itemize}
    \item Channel Collector \& Predictor: Fetches raw CSI and automatically triggers the NNCF algorithm if data missing is detected.
    \item Optimization Solvers: Three distinct Python-based solvers corresponding to the optimization problems P1, P2, and P3.
\end{itemize}
This modular design ensures that the agent can dynamically chose the specific tool required for the current context.

\subsubsection{Orchestration Layer (Workflow)}
The Coze platform manages the ``Sense-Repair-Decide-Act'' data pipeline. It employs a data-unifier method to standardize inputs from different sources (raw or repaired CSI) and routes them to the appropriate solver based on the recognized intent, which reduces the operational complexity and delay of resource scheduling.

\section{Algorithm Design}
\label{sec:algorithm}

The proposed optimization problems in Section \ref{sec:system_model} are non-convex and involve coupled integer variables (subcarrier allocation) and continuous variables (power allocation). To solve these problems, we propose a two-step framework. In the first step, we recover the incomplete channel matrix using an AI-based method. Subsequently, we employ specific resource allocation algorithms according to the user's intent.

\subsection{AI-Based Channel Reconstruction}
Let $\mathbf{G}_{raw}$ be the incomplete channel matrix acquired by the BS. We propose a NNCF algorithm to estimate the missing CSI. This method relies on the correlation of channel gains in both the frequency domain and spatial domain.

For a missing entry at index $(k, n)$, we define a neighbor set $\mathcal{N}_{k,n} = \{(k, n-1), (k, n+1), (k-1, n), (k+1, n)\}$. The reconstructed value $\hat{g}_{k,n}$ is computed as the weighted average of valid neighbors:
\begin{equation}
    \hat{g}_{k,n} = \frac{\sum_{(i,j) \in \mathcal{N}_{k,n} \cap \Omega} w_{i,j} \cdot g_{i,j}}{\sum_{(i,j) \in \mathcal{N}_{k,n} \cap \Omega} w_{i,j}},
\end{equation}
where $w_{i,j}$\cite{9245325,7349760}represents the correlation weight. We assign higher weights to frequency-domain neighbors due to the usually higher correlation across near subcarriers. The algorithm cycleses until all missing datas are filled, making the complete matrix $\hat{\mathbf{G}}$.

\subsection{Intent-Driven Resource Scheduling Algorithms}
Based on the recognized user intent, the system selects one of the following algorithms to optimize resource allocation using $\hat{\mathbf{G}}$.

\subsubsection{Algorithm for Max Rate (Water-filling)}
For Problem P1, we decouple the subcarrier and power allocation. To maximize sum rate, each subcarrier $n$ is assigned to the user with the best channel gain:
\begin{equation}
    x_{k,n}^* = \begin{cases} 
    1, & \text{if } k = \arg\max_{i \in \mathcal{K}} \hat{g}_{i,n} \\
    0, & \text{otherwise}.
    \end{cases}
\end{equation}
With fixed $x_{k,n}^*$, the problem becomes convex. Using the Lagrangian method, the optimal power allocation follows the water-filling principle:
\begin{equation}
    p_{k,n}^* = x_{k,n}^* \left[ \frac{1}{\lambda} - \frac{N_0 B}{\hat{g}_{k,n}} \right]^+,
\end{equation}
where $[z]^+ = \max(0, z)$, and $\lambda$ is the Lagrange multiplier determined by the total power constraint $\sum p_{k,n} = P_{max}$ via bisection search.

\subsubsection{Algorithm for Min Power }
For Problem P2, the goal is to satisfy the QoS constraint $R_{min}$ with minimal power.
We employ a heuristic approach where each user $k$ is pre-assigned a set of subcarriers $\mathcal{S}_k$ with the highest gains for that user, ensuring fairness.
To minimize power, we apply channel inversion. Assuming the rate requirement is equally distributed among assigned subcarriers, the required power is:
\begin{equation}
    p_{k,n}^* = \frac{(2^{R_{min} / (B |\mathcal{S}_k|)} - 1) N_0 B}{\hat{g}_{k,n}}, \quad \forall n \in \mathcal{S}_k.
\end{equation}

\subsubsection{Algorithm for Max EE }
Problem P3 is a fractional programming problem. Since the EE function is generally quasiconcave with respect to the total transmit power $P_{total}$, we propose a low-complexity iterative search algorithm.
We discretize the power range $(0, P_{max}]$ into $L$ levels. For each power level $P_{l}$, we solve the max rate problem (as in Scenario 1) with $P_{l}$ as the power budget to obtain the maximum rate $R_{sum}(P_{l})$. The energy efficiency is then calculated as:
\begin{equation}
    \eta(P_{l}) = \frac{R_{sum}(P_{l})}{P_{l} + P_c}.
\end{equation}
The optimal power allocation corresponds to the level $P_{l}^*$ that maximizes $\eta(P_{l})$.

\subsection{Overall Algorithm Procedure}
The complete execution flow of the proposed framework is summarized in Algorithm 1. The computational complexity of the proposed framework is dominated by the iterative processes. Specifically, the NNCF channel repair requires $\mathcal{O}(I_{iter}KN)$ operations. For resource scheduling, the complexity is upper-bounded by the Max EE mode at $\mathcal{O}(LNK)$, where $L$ is the number of power quantization levels. This linear scaling with respect to $N$ ensures the algorithm's feasibility for real-time implementation.

\begin{algorithm}[!t]
\caption{AI-Enabled Resilient Resource Allocation Framework}
\label{alg:overall}
\begin{algorithmic}[1]
\REQUIRE Raw CSI $\mathbf{G}_{raw}$, User Intent $I$, Constraints $P_{max}, R_{min}, P_c$.
\ENSURE Optimal Subcarrier $\mathbf{X}^*$ and Power $\mathbf{P}^*$.

\STATE \textbf{Phase 1: AI-Based Channel Reconstruction}
\STATE Identify missing entries (indices with NaN) in $\mathbf{G}_{raw}$.
\STATE Reconstruct complete matrix $\hat{\mathbf{G}}$ using NNCF algorithm by exploiting frequency-spatial correlations.

\STATE \textbf{Phase 2: Intent-Driven Resource Scheduling}
\IF{$I = \text{Max Rate}$}
    \STATE Perform subcarrier allocation $\mathbf{X}^*$ via greedy strategy based on channel gains.
    \STATE Calculate power $\mathbf{P}^*$ via \textbf{Water-filling} algorithm under budget $P_{max}$.
\ELSIF{$I = \text{Min Power}$}
    \STATE Allocate subcarriers to meet minimal QoS requirement.
    \STATE Calculate power $\mathbf{P}^*$ via \textbf{Channel Inversion} to satisfy $R_{min}$ with minimal energy.
\ELSIF{$I = \text{Max EE}$}
    \STATE Initialize optimal efficiency $\eta^* = 0$.
    \FOR{each power level $P_l \in (0, P_{max}]$}
        \STATE Solve Max Rate problem with temporary budget $P_l$.
        \STATE Calculate efficiency $\eta_l = R_{sum}(P_l) / (P_l + P_c)$.
        \IF{$\eta_l > \eta^*$}
            \STATE Update $\eta^* \leftarrow \eta_l$, $\mathbf{X}^* \leftarrow \mathbf{X}_l$, $\mathbf{P}^* \leftarrow \mathbf{P}_l$.
        \ENDIF
    \ENDFOR
\ENDIF

\STATE \textbf{Phase 3: Execution and Configuration}
\STATE Output optimal resource allocation schemes $\mathbf{X}^*$ and $\mathbf{P}^*$ to the Base Station.
\end{algorithmic}
\end{algorithm}

\section{Simulation Results}
\label{sec:results}

\subsection{Simulation Setup}
To evaluate the performance of the proposed AI-enabled resource allocation framework, we conduct simulations considering a downlink OFDMA system. The simulation parameters are configured as follows: the system serves $K=8$ users with $N=20$ subcarriers. The subcarrier bandwidth $B$ is set to $180$ kHz, and the noise power spectral density $N_0$ is $1 \times 10^{-17}$ W/Hz. The maximum transmit power of the BS is $P_{max} = 20$ W. The circuit power consumption $P_c$ is set to $5$ W.

To model the imperfect CSI case, we generate a frequency-selective fading channel matrix based on the Rayleigh fading model\cite{9921202}, incorporating path loss components. We assume a random data loss rate of $20\%$, where entries in the raw channel matrix $\mathbf{G}_{raw}$ are randomly masked as missing values to simulate environmental interference or feedback failures.

\subsection{Channel Reconstruction Performance}
We check the effectiveness of the proposed AI-based channel reconstruction module NNCF by comparing it with a traditional statistical imputation method (filling missing values with the user's average channel gain).

\begin{figure}[!t]
\centering
\includegraphics[width=\columnwidth, trim=0.0cm 0.0cm 0.0cm 0.0cm, clip]{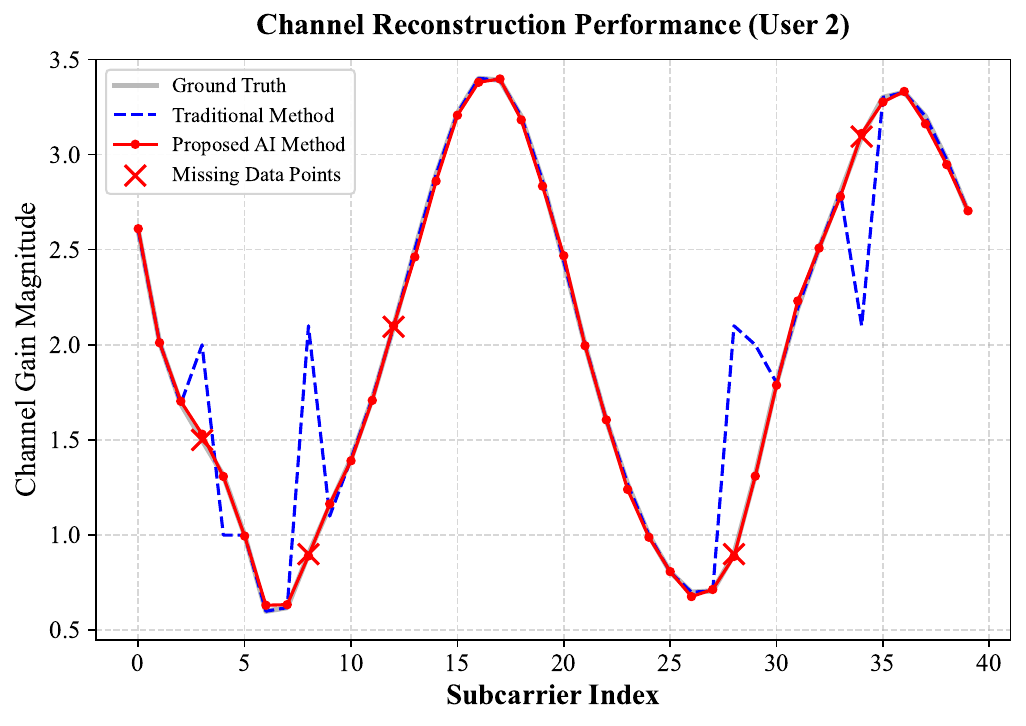}
\caption{Comparison of channel gain reconstruction for a randomly selected user. The black solid line represents the ground truth, the blue dashed line represents the traditional mean imputation, and the red marker line represents the proposed AI-based repair method.}
\label{fig:channel_recovery}
\end{figure}

Fig. \ref{fig:channel_recovery} illustrates the channel recovery performance across subcarriers for a specific user. As observed, the actual wireless channel (Ground Truth) exhibits frequency-selective fading characteristics, with peaks and deep valleys. The traditional method simply fills missing points with a constant average value, failing to capture the dynamic fluctuations of the channel. Consequently, this leads to estimation errors, particularly in deep fading or high gain regions. In contrast, our proposed AI method exploits the strong correlation between adjacent subcarriers in the frequency domain. It accurately reconstructs the missing channel gains, fitting the trend of the ground truth curve closely. This demonstrates that the NNCF algorithm effectively restores the detailed features of the CSI.

\begin{figure}[!t]
\centering
\includegraphics[width=\columnwidth, trim=0.0cm 0.0cm 0.0cm 0.0cm]{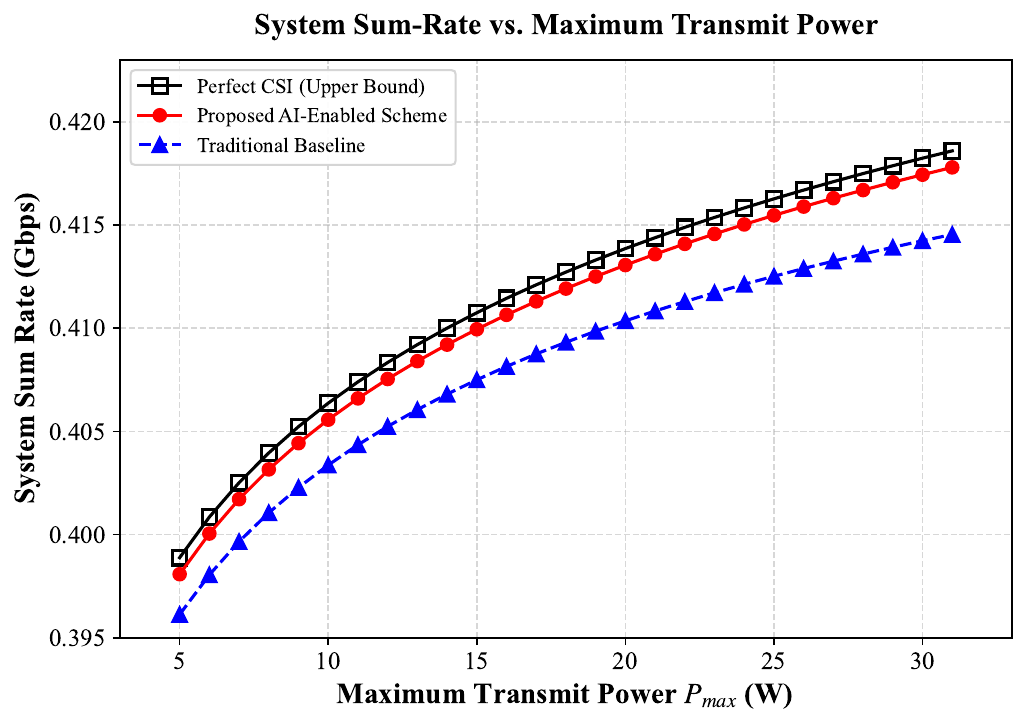}
\caption{System sum rate versus maximum transmit power $P_{max}$ under $30\%$ channel data loss.}
\label{fig:sum rate}
\end{figure}

Fig. \ref{fig:sum rate} illustrates the system sum-rate performance as a function of the maximum transmit power budget $P_{max}$. As observed, the proposed AI-enabled scheme (Red line) achieves a performance close to the upper bound with Perfect CSI (Black line), maintaining a small gap across the entire power range. In contrast, the traditional baseline (Blue dashed line) suffers from noticeable performance degradation due to inaccurate channel estimation. These results confirm that the high-precision channel reconstruction provided by the AI module directly translates into significant system-level throughput gains.

\subsection{Resource Allocation Performance}
Based on the repaired CSI, the system executes intent-driven resource scheduling. 
In the Max Rate mode, the system allocates the full power budget ($20$ W) to subcarriers with high channel gains, maximizing the spectral efficiency. 
In the Min Power mode, the system successfully calculates the minimum power required to meet the QoS constraints, significantly reducing energy consumption compared to fixed-power schemes. 
In the Max EE mode, the algorithm identifies the optimal operating point where the ratio of throughput to power consumption is maximized, demonstrating the system's capability to balance performance and sustainability adaptively.

\section{Conclusion}
\label{sec:conclusion}

In this paper, we proposed an AI-enabled resource allocation framework for OFDMA systems operating under partial CSI. We addressed the challenge of CSI loss by integrating a NNCF algorithm, which effectively reconstructs missing channel data by exploiting intrinsic frequency-spatial correlations. Furthermore, we formulated a multi-objective optimization problem that intelligently adapts to diverse user intents, enabling the system to switch seamlessly between maximum rate, minimum power, and maximum EE strategies.

To solve the formulated problems, we developed a series of efficient algorithms, including water-filling, channel inversion, and iterative power search, based on the reconstructed CSI. Simulation results validated that our proposed AI-based repair method significantly reduces channel estimation errors compared to traditional approaches. Moreover, the intent-driven scheduling system was shown to achieve superior system performance, flexibly balancing high throughput, low power consumption, and energy efficiency. Future research could extend this framework to multi-cell massive MIMO scenarios, incorporate deep reinforcement learning for real-time intent recognition, and explore robustness under high-mobility conditions.

\bibliographystyle{IEEEtran}  
\bibliography{ref}            
\end{document}